\documentclass[preprint,review,12pt]{elsarticle}
\usepackage{graphicx}
\usepackage{xcolor}
\usepackage[normalem]{ulem}
\usepackage[utf8]{inputenc}
\usepackage{amsmath}
\usepackage{amssymb}
\journal{Physica E}
\begin{document}

\begin{frontmatter}

\title{Evidence for flat zero-energy bands in  bilayer graphene with a periodic defect lattice}

\author{A. K. M. Pinto}
\address{Departamento de Física,  Universidade Federal Rural de Pernambuco, Rua Dom Manoel de Medeiros s/n, Dois Irmãos, 52171-900, Recife, PE, Brazil} 
\address{ and Departamento de Física, Universidade Federal da Paraíba, Caixa Postal 5008, 58059-900, João Pessoa, PB, Brazil.}
\ead{munizallan@gmail.com}

\author{N. F. Frazão}
\address{PPGCNBiotec, Centro de Educação e Saúde, Universidade Federal de Campina Grande, Campus Cuité, 58175-000, Cuité, PB, Brazil
} 
\address{and Departamento de Física, Universidade Federal de Campina Grande, Caixa Postal 10071, 58109-970, Campina Grande, PB, Brazil
}
\ead{nilton.frazao@ufcg.edu.br}

\author{D. L. Azevedo}
\address{Instituto de Física, Universidade de Brasília, Campus Universitário Darcy Ribeiro, Asa Norte, 70919-970, Brasília, DF, Brazil
}\address{and Faculdade UnB Planaltina, Universidade de Brasília, Área Universitária 01, 73345-010, Planaltina, DF, Brazil}
\ead{david-azv@fis.unb.br}

\author{F. Moraes}
\address{Departamento de Física, Universidade Federal Rural de Pernambuco, Rua Dom Manoel de Medeiros s/n, Dois Irmãos, 52171-900, Recife, PE, Brazil}
\ead{fernando.jsmoraes@ufrpe.br}

\vspace{10pt}

\begin{abstract}
In this work, we perform \textit{ab initio} calculations, based on the density functional theory, of the electronic structure of a graphene bilayer with a periodic array of topological defects. The defects are generated by  intercalation of carbon atoms between the layers and their subsequent absorption by one of the layers. We use the unit cell of the bilayer to construct larger unit cells (supercells),  positioning a single carbon atom in the \textit{hollow} position between the monolayers and    periodically replicating the supercell. By increasing the size of the supercell and, consequently, the periodicity of the inserted atoms,  we not only optimize the results but also vary the size of the defect lattice. Our main result  is the appearance of a doubly degenerate flat band at the Fermi level. These states are interpreted as coming from the periodic deformation of the bilayer due to the topological defects generated by the inserted atoms. It acts as a non-Abelian flux network creating zero energy flat bands as predicted by San-Jose, González and Guinea in 2012. {Since the periodic strain field  associated to the  defect array has such a strong influence on the electronic properties of the bilayer, it may be useful for practical applications. For instance, it can act as frozen-in magnetic-like field flux tubes. All-carbon nanostructures can then be designed to have   electronic behavior at different regions tailored by the chosen defect  pattern. }

\end{abstract}

\end{frontmatter}

\section{Introduction}

The advent of carbon-based low-dimensional materials, like nanotubes, fullerenes, single layer graphene (SLG) and bilayer graphene (BLG), to name a few, brought forward very rich physics and a wide range of applications. While SLG has a zero band gap and linear dispersion relation, allowing a description by massless Dirac fermions \cite{novoselov2005two},   BLG has an easily tunable electronic band gap and parabolic dispersion relation, when in the A-B,   (or Bernal-stacked) form \cite{mccann2006asymmetry}. These differences make BLG more attractive to electronic and photonic applications, without losing its fundamental physics appeal, as illustrated by the recently reported discovery of the exotic 5/2 quantum Hall state in BLG \cite{li2017even} and unconventional supercondutivity in twisted BLG \cite{cao2018unconventional}. Moreover, electrons in SLG (BLG) are chiral quasiparticles which, when adiabatically propagated along a closed orbit, acquire a  Berry phase of $\pi$ ($2\pi$), according to their degree of chirality  \cite{mccann2006landau}. 

{It is well known that mono and bilayer graphene and graphene-based materials have found their way to many technological applications \cite{corocs2019brief}. Nevertheless, other two-dimensional systems like hexagonal boron-nitride (h-BN) \cite{brito2019theoretical}, phosphorene \cite{le2019combined,le2019perpendicular,le2019perturbation}, borophene \cite{le2019beta}, silicene \cite{zhao2016rise},  germanium monosulfide \cite{zhang2016two,le2019strain}, siligraphene \cite{le2019real,hussain2019three}, etc., also appear as promising materials for  nanoelectronics, spintronics and optoelectronics applications.}

The recently reported superconductivity in twisted BLG \cite{cao2018unconventional,cao2018correlated} turns it even more important for the study of fundamental physics as well as for applications. Theory has predicted that flat bands, not only enable superconductivity but also can enhance its transition temperature \cite{kopnin2011high}. Flat bands in BLG  could be obtained by doping or gating but, in twisted bilayer graphene (TBLG), they arise as a consequence of the moir\'e pattern associated to the twist \cite{san2012non}. As explained in  reference \cite{san2012non}, the moir\'e superlattice appears to the electrons in TBLG as a periodic array of non-Abelian gauge fluxes, which leads to localized Landau-like levels, the flat bands. These flat bands start to form as the pattern period increases. In this work, by doing DFT calculations, we show that BLG periodically intercalated with carbon atoms  presents a similar behavior. The inserted atoms bond to one of the layers producing a periodic array of topological defects. We observe localized bands near the defects, which gradually flatten out as the separation between the defects increases.  The local deformation of the BLG lattice due to each intercalated atom acts as a pseudomagnetic flux \cite{manes2007symmetry,guinea2010energy,levy2010strain} giving rise to the observed localized states. This way, we present evidence that  a periodic array of topological defects generated by intercalation of carbon atoms in BLG leads to flat bands as observed in TBLG. {An exciting experimental perspective is to combine twisting with the introduction of topological defects, in contrast with twisting plus doping done in \cite{cress2016nitrogen}.}

{Reference \cite{liu2018bubble} reports an extensive study of periodic arrays of fullerenes incorporated into a single layer graphene sheet, so-called bubble-wrap carbon. The electronic structure there obtained shows clearly the emergence of flat bands. Also, metallic behavior is achieved at high density of the bubbles. Although both SLG and BLG are very important for technological applications, since their basic physics are quite different, it is important to verify if anything similar to what was observed in \cite{liu2018bubble} also happens in BLG. For this end we choose to work with BLG with the simplest ``bubble'' we could think of, a topological defect formed by the periodic absorption of a carbon atom by one of the layers (see Fig. \ref{estruturas}). } We focus on the most stable graphene bilayer, which consists of two graphene monolayers stacked in a model known as A-B Bernal phase, in which one of the layers is rotated by $180^{\circ}$ relative to one another and the interaction between the layers is of van der Waals type \cite{mccann2006landau}. For  recent reviews on electronic properties of this and the other types of BLG see Refs. \cite{mccann2013electronic} and \cite{rozhkov2016electronic}.

\section{Computational Methodology}

 The graphene bilayer can be obtained experimentally by Chemical Vapor Deposition \cite{yan2011formation} and its lattice parameters  are $a=b= 2.456$~{\AA}   ($a$ and $b$ indicate the crystal lattice constants in the $xy$ plane) \cite{slonczewski1958band}. Its unit cell has four atoms   and the separation distance between the monolayers is $3.34$~{\AA}  \cite{choi2016graphene}. Using these data, we performed  first principles calculations of the graphene bilayer intercalated with  carbon atoms in the hollow position. The calculations were done using the CASTEP code, which is included in the Materials Studio software package and uses the density functional theory plane-wave pseudopotential method   (DFT) \cite{hohenberg1964p,kohn1965w}. The CASTEP code also  uses  the generalized gradient approximation (GGA), and the  Perdew-Burke-Ernzerhof (PBE) functional  \cite{perdew1996generalized}.
 
In our calculations  we used norm-conserved pseudopotential, which increases the computational cost, but yields good results. The Monkhorst-Pack grid \cite{PhysRevB.13.5188} was 11$\times$11$\times$1, for a better analysis of integrals in the reciprocal space. The structures used were optimized with the total minimum energy, in which the maximum energy change during geometry optimization was $0.001$~{eV/cell}, the maximum force is $0.05$~{eV/\AA}, pressure smaller than  $ 0.01$~{GPa}, and maximum atomic displacement not exceeding $0.002$~{\AA}. Furthermore, we had a self-consistent field  (SCF) tolerance of 1$\times10^{-4}$~{eV/cell}, and convergence window of three cycles. The energy cutoff is $680$~{eV}.

\section{Results and discussion}

\subsection{Geometry optimization}

\begin{figure}
 \centering
 \includegraphics[scale=0.60]{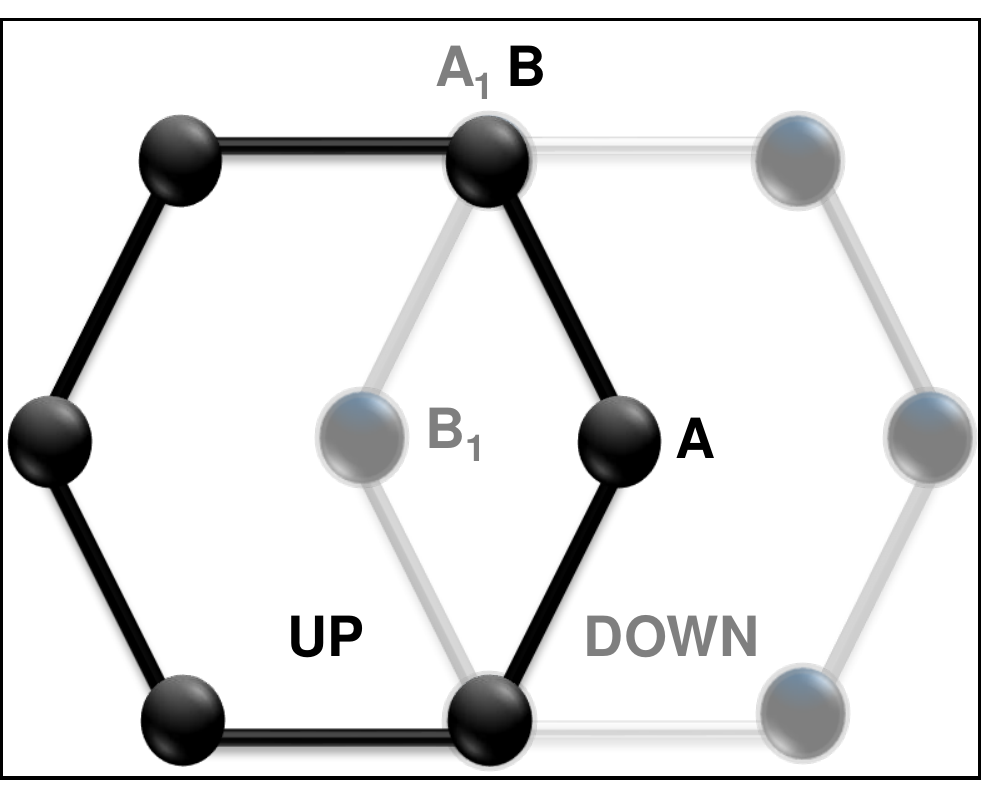}
 \caption{\label{empilhamento} Bernal stacking scheme for the graphene bilayer. In black we have the upper hexagon and,  in gray, the lower hexagon.}
\end{figure}

The top view of the Bernal stacking of our system is shown in Fig. \ref{empilhamento} where gray represents the lower layer and black represents the upper layer. As we can see, position $A$ of the top layer lies exactly above the center of the hexagon of the bottom layer. The $B$ position of the upper layer coincides with the $A_{1}$ position of the lower layer and the $B_{1}$ position of the lower layer exactly matches the center of the hexagon of the upper layer. Finally, in the hollow position, the atom is above the center of the hexagon formed by the lower monolayer, as shown in  Figure \ref{empilhamento}.

\begin{figure*}[htb!]
 \centering
 \includegraphics[scale=0.27]{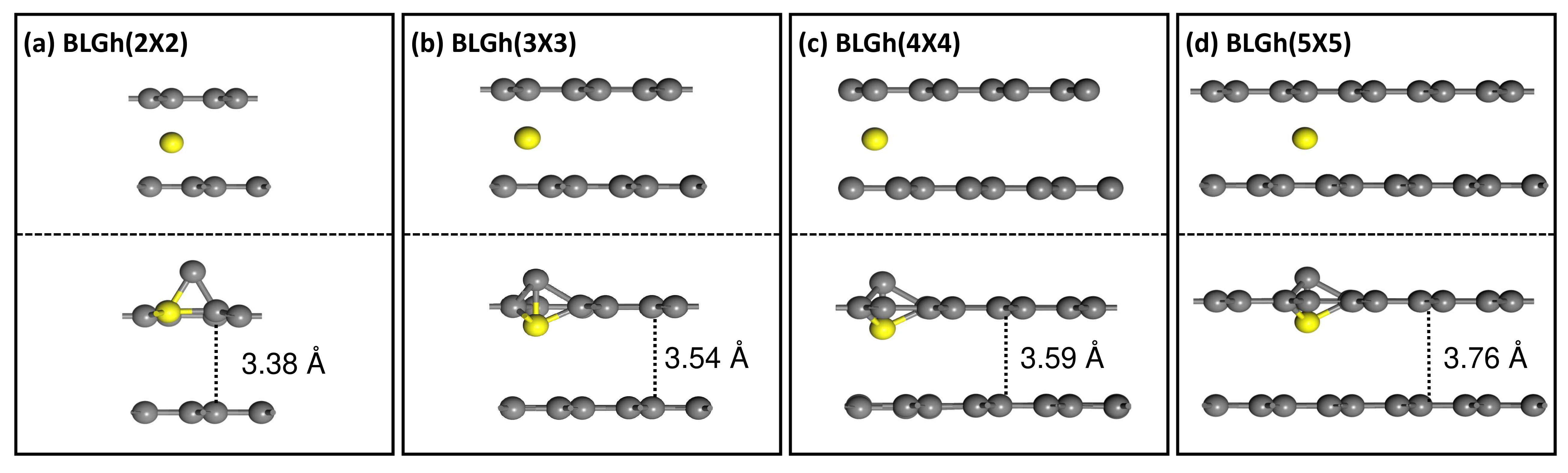}
 \caption{\label{conver}  Upper figures: representation of the intercalated carbon atom for the different supercell sizes before optimization. Lower figures: topological defects formed upon optimization.  }
 \label{estruturas}
\end{figure*}

\begin{figure}[htb!]
 \centering
 \includegraphics[scale=0.40]{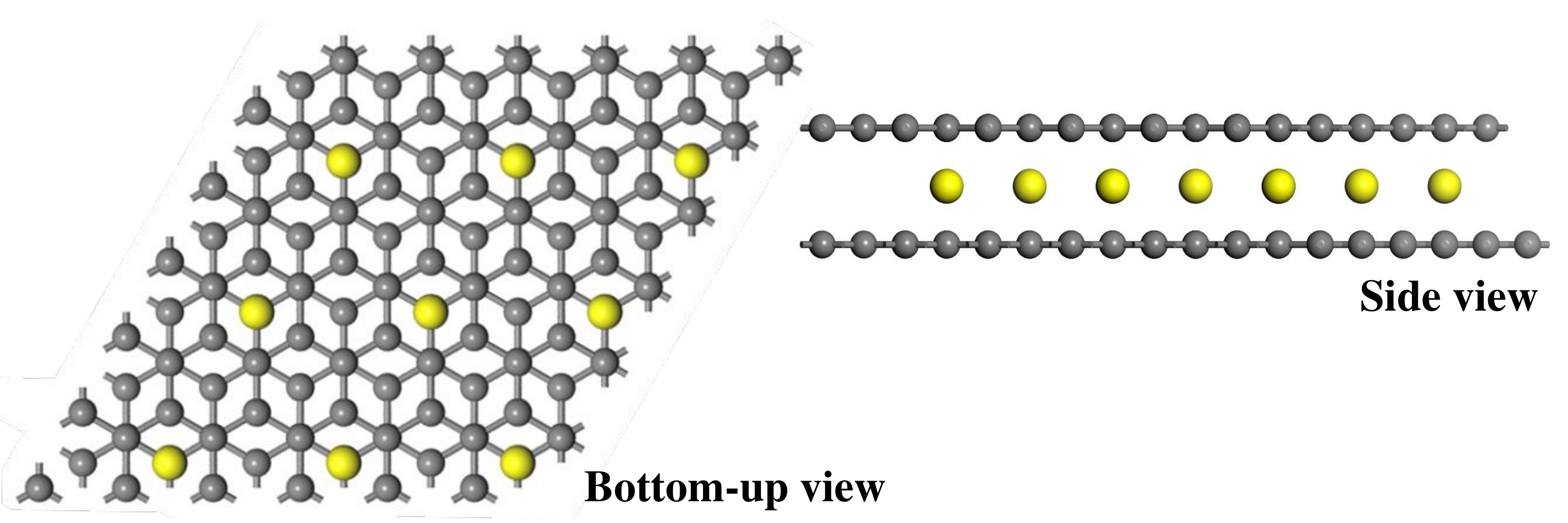}
 \caption{Representation of the carbon-intercalated BLG obtained from the repetition of the 2$\times$2 supercell before optimization. \label{structview}}  
\end{figure}

In our study, we used the supercell scheme, in which the cells are periodically arranged in space.  We used 4 different sizes for the supercell: 2$\times$2, 3$\times$3, 4$\times$4  and 5$\times$5, each with 16, 36, 64 and 100 carbon atoms, respectively (not counting the extra atom).  %we constructed a group of four structures to represent the pure graphene bilayer. They are labeled BLGp(2X2) (16 atoms), BLGp(3X3) (36 atoms), BLGp(4X4) (64 atoms) and BLGp(5X5) (100 atoms). The other group of structures, with the C atom inserted in the hollow position, are BLGh(2X2) (17 atoms), BLGh(3X3) (37 atoms), BLGh(4X4) (65 atoms) e BLGh(5X5) (101 atoms).  
The four structures are represented by their supercells in Fig. \ref{estruturas}. The lattice parameters of each structure is shown in Table \ref{table1}. With respect to the lattice parameter $c$, associated to the periodicity  in the direction normal to the bilayers, we always used  $c = 20$~{\AA}  to ensure that there is no interaction between the multiple bilayers that  arise  due to the periodicity of the system.

In the optimization geometry, our structures were relaxed until they reached the convergence criteria. As can be seen in all cases depicted in Fig. \ref{conver},  the inserted atom binds to the upper layer pushing one of its atoms to an out-of-plane position. %where in the case of graphene bilayers with the intercalated carbon atom, inserted atom binds the upper monolayer to the BLG(2X2) system, the inserted $C$ atom it displaces one of the atoms and assumes its position in the upper monolayer causing a deformation in that region, as can be seen in figure \ref{conver} (a)(down), already in the BLG(3X3) system, for example, 
Thus, when the C atom joins the monolayer, it does not only assume the position of one of the atoms, but rather   reorganizes the bonds in that region, deforming the structure. In fact, as it is clear form Fig. \ref{estruturas}, the three neighboring atoms  of the one above the intercalated atom change their hybridization from  $sp^2$ to $sp^3$. Consequently,  the intercalated atom and its upstairs neighbor are pushed out of the graphene plane. The resulting structure cannot be relaxed back to the hexagonal lattice without cutting and rejoining chemical bonds, what characterizes a topological defect.  The  repetition  of the supercell with the extra atom in 2D space then generates an infinite BLG with a periodic  deformation reminiscent of bubble-wrap graphene \cite{liu2018bubble}. In Fig. \ref{structview} it is shown a representation of the lattice generated by the 2$\times$2 supercell before optimization.

In  Table \ref{table1} we have the superlattice parameter $L$, which corresponds to the periodicity of the defect array, for each of the structures studied. $L$ was obtained from the  value for the pristine BLG unit cell, $a,b=2.46$~{\AA},  by appropriate scaling. The parameter $c = 20$~{\AA} is the same for all structures.   $dE$ is the binding energy of the system, which is obtained from the expression,

\begin{equation}
\centering
 dE = E_{BLGp} + E_{atom} - E_{BLGh},
\end{equation}
where $E_{BLGp}$ is the energy of the pure bilayer, $E_{atom}$ the energy of a single C atom, and $E_{BLGh} $ is the energy of the system with the atom intercalated in the bilayer. $a_{C-layer}$ corresponds to the  distance between the intercalated C atom and  its  closest neighbor in the monolayer. $S_{A-B}$  is the distance between the monolayers after the atom is added to the system, whose value for pure bilayers is $3.34$~{\AA} \cite{choi2016graphene}.
  
We observe that the formation energy of BLG(3$\times$3), BLG(4$\times$4) and BLG(5$\times$5) scale approximately with the cell size. The higher value for BLG(2$\times$2) seems to be due to the interaction between defects in neighboring cells. This effect naturally dies off quickly for larger cells.
%the formation energy does not follow an increasing order according to the size of the supercell, once higher energy of formation was BLG(2X2), with $1.41$~{eV}; already BLG(3X3) presents a value of $0.54$~{eV}; in the sequence we have BLG(4X4), with $ 0.65$~{eV}; and BLG(5X5) with $ 1.04$~{eV} . So we have the energy system formation does not have any relation to its scale. 
In all systems investigated, the bond distances between the intercalated atom and the monolayer atom  attached to it are very close to the usual $C-C$ single bond distance of $ 1.54$~{\AA}   \cite{atkins2016chemical}.  %in the monolayer is $ 1.54$~{\AA}  in the BLG(2X2) system and from $ 1.59$~{\AA}  to BLG(3X3), the others got $ 1.58$~{\AA}, which are acceptable distances, since the usual $C-C$ single bond distance is $ 1.53$~{\AA}   \cite{rocco}.
The  separation of the layers near the defects follows an increasing order as the defect lattice parameter increase, from the lower value $3.38$~{\AA} for the BLG(2$\times$2) system, to the higher  value $ 3.76$~{\AA}  for BLG(5$\times$5). Away from the defects the separation is $ S_{A-B} = 3.34$~{\AA}, which is the value for the pure bilayer. Throughout the text we distinguish the pristine and hollow-intercalated structures by referring, respectively, to BLGp and BLGh.

\begin{table}
\caption{\label{table1} The periodicity of the topological defect array ($L$), binding energy ($dE$), distance between adsorbed atom  and its monolayer ($a_{C-layer}$), and spacing between  the monolayers ($S_{A-B}$).}
\footnotesize\rm
\begin{tabular}{l l l l l }
%\br
%\mr
\hline
  & BLG(2$\times$2)&BLG(3$\times$3)&BLG(4$\times$4)&BLG(5$\times$5)\\
%\mr
\hline
$L$ (\AA)&4.92& 7.38&9.84&12.30\\
$dE$ (eV)&1.41&0.54&0.65&1.04\\
a$_{C-layer}$ (\AA)&1.54&1.59&1.58&1.58\\
\textit{$S_{A-B}$} (\AA)&3.38&3.54&3.59&3.76\\
\hline
%\br
\end{tabular}
\end{table}

For the sake of verification of mechanical and thermodynamical stability of the structures, we  also calculated the  phonon spectrum, phonon density of states and the thermodynamic functions (enthalpy, entropy, Gibbs free energy, heat capacity and Debye temperature), which will be the subject of a separate publication.  We observed that all structures studied are stable, since they present only positive frequencies and  can be built spontaneously (an exothermic process). 

\subsection{Flat zero-energy bands}

\begin{figure}[htb!]
 \centering
 \includegraphics[scale=0.25]{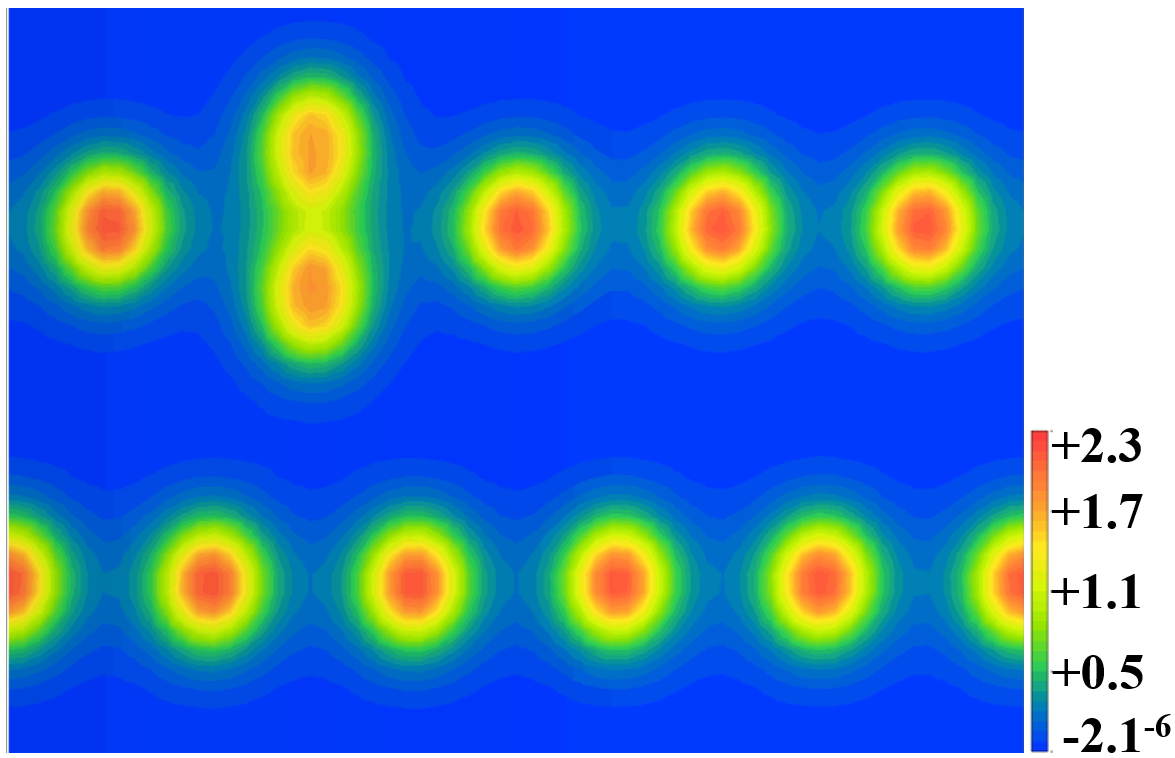}
 \caption{Charge density profile of a section of  the  BLGh(5$\times$5) supercell with the topological defect. Neither the extra nor the dislocated atom are  on the plane of the cut. \label{densityblg5x5}}  
\end{figure}

The fact that the intercalated atoms get incorporated into one of the graphene layers without really doping it (not transferring charge) generates a structurally stable lattice of local deformations. For a single graphene layer these deformations are well described \cite{manes2007symmetry} by an Abelian gauge field; i.e., a pseudomagnetic field. This gives rise to localized, Landau-like, levels \cite{guinea2010energy}. In the case of the bilayer, the interlayer hopping is affected by the distortions in such a way that the effective field is now non-Abelian \cite{san2012non}. In this reference, it is shown that a periodic pattern, with large periodicity, of such non-Abelian fluxes induces spatial confinement of electronic states in flat bands. As the periodicity of the pattern increases they observed that ``the system develops two increasingly narrow sub-bands around zero energy states''. This is exactly what we obtain in our study of periodically C-intercalated BLG, as shown in Fig. \ref{eletronic}.  Even though the results of Ref. \cite{san2012non} were obtained for sheared or twisted bilayers, in the  resulting structure, the interlayer separation is periodically corrugated as it also happens in our system. This is clear in Fig. \ref{densityblg5x5} where we show the charge density profile in a transverse cut of our system.

\subsection{Band Structure and Density of States}

\begin{figure*}[htb!]
 \centering
 \includegraphics[scale=0.18]{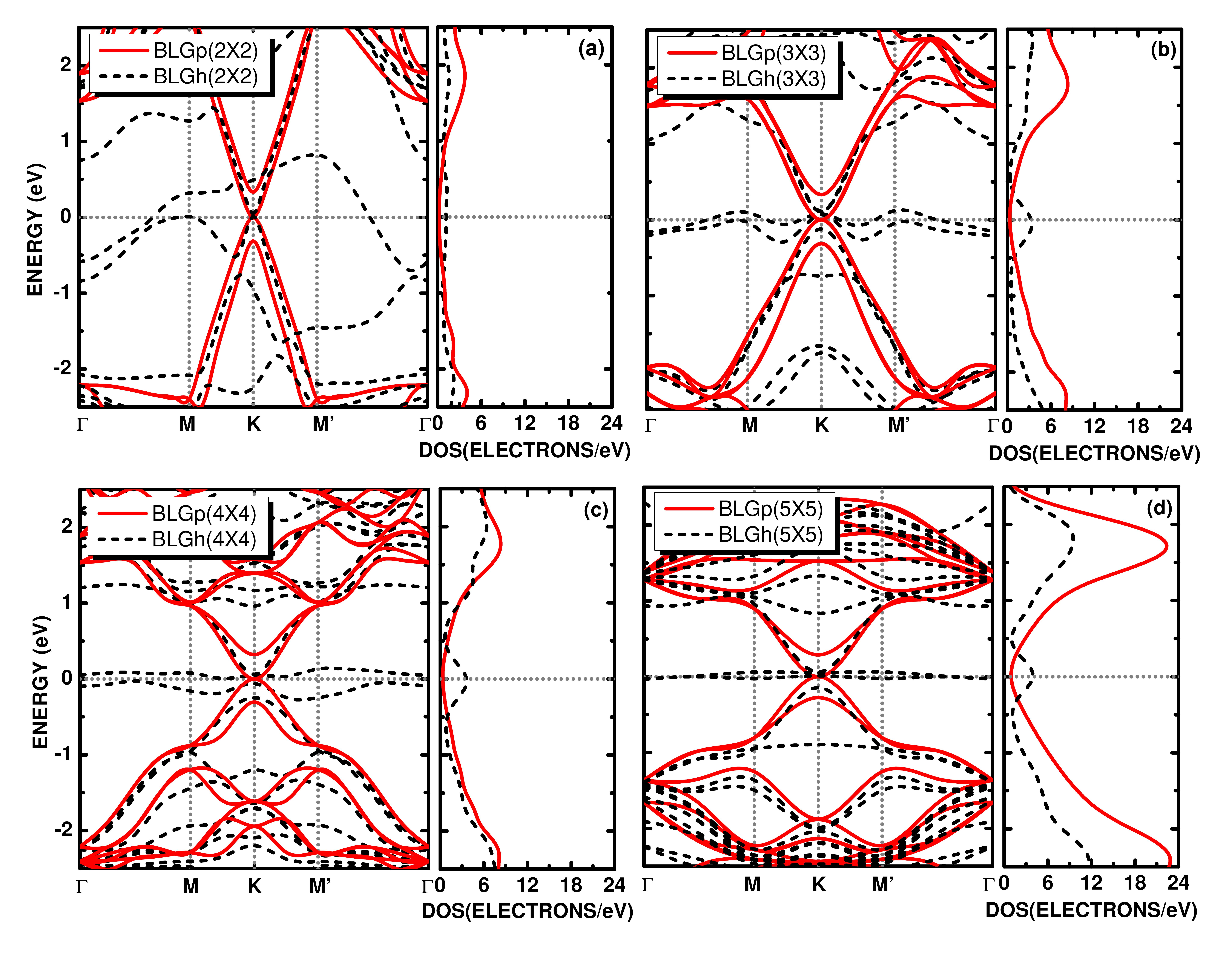}
 \caption{\label{figureone} The band structures and density of states of the graphene bilayer is shown for the  (a) BLG(2$\times$2), (b) BLG(3$\times$3), (c) BLG(4$\times$4), and (d) BLG(5$\times$5) supercells. The solid red lines represent the pure bilayer graphene system, the dotted black lines correspond to the bilayer graphene systems with topological defects created by the extra C atoms. The Fermi energy is  the dotted grey line at zero energy.}\label{eletronic}
\end{figure*}

We now present the calculated energy bands as a function of the periodicity of the defect lattice (or supercell size). The analysis is performed by observing the bands in relation to the points of high symmetry in the first Brillouin zone. We also compute both the total and the spin-resolved  density of states (DOS) for each case.

The high symmetry points are in the following sequence: $\Gamma$ (0.000, 0.000, 0.000), M (0.000, 0.500, 0.000), K (0.333, 0.333, 0.000), M’ (0.500, 0.000, 0.000) and $\Gamma$ (0.000, 0.000, 0.000). In Fig. \ref{eletronic}, we present the calculated band structure and corresponding  density of states for each supercell size. For the sake of comparison, each graph includes  both   BLGh (dotted black lines) and BLGp (red lines). %We also have points of high symmetry (vertical dotted gray line) and the Fermi level (horizontal dotted  gray line). Still in figure \ref{eletronic}, we have the axis of the energy that goes from $ -4$~{eV} to $4$~{eV} , being that the Fermi Energy level is set to be zero.

Ideally, the calculations should be performed for the actual experimental size of the bilayer, which is unfeasible due to the large number of atoms involved. A way of getting by this difficulty is by defining a unit cell and then using periodic boundary conditions. Obviously, one expects that the larger the unit cell the more realistic the results will be.  For this matter, we did calculations for four different supercell sizes in order to infer any tendency of convergence of the computed physical observables. Besides improving the quality and reliability of the results, this procedure naturally gives us a way of changing the lattice parameter of the topological defect array. In figure \ref{eletronic} we show the band structure and density of states for the four cases,  both with and without the extra carbon atoms. In the pristine bilayer (BLGp) case, the usual gapless parabolic dispersion relation appears clearly for all supercell sizes, in agreement with the literature \cite{ohta2006controlling}. For the intercalated BLGh a small band gap appears at the K point which tends to $\approx$ 0.2 eV as one goes to the larger structures. Further, starting at the 3$\times$3 structure, there appears two energy bands  spread around the Fermi level. As the size of the structure increases, those bands  converge to a single flat band at the Fermi level. A doubly degenerate zero mode which is clearly seen in the density of states plots of the corresponding structures.

%We note that in BLG(2$\times$2), figure \ref{eletronic}~{(a)}, for example, BLGp(2$\times$2) has a parabolic energy dispersion, where the bands touch the point $K$, relevant behavior with what we have in the literature \cite{ohta2006controlling}, the same applies to the BLGp(4$\times$4) and BLGp(5$\times$5) systems, figure \ref{eletronic} (c) and (d). However, the BLGp(3$\times$3) system, bands prefer to touch the Gamma point, figure \ref{eletronic}~{(b)}. For us, We put the atom free to building bond interaction, for example: in the BLG(3$\times$3) structure the atom chose a path way different, in comparison with the other bilayers. This results a remarkable difference in the band structure plot (see figure \ref{eletronic}~{(b)}).

%In the DOS of the BLGp, for all systems on the Fermi level we have a minimum point, showing that we do not have gap energy, i.e., on the energy of Fermi no state can be occupied, which is in agreement with the literature \cite{ohta2006controlling}.

\begin{figure*}[htb!]
 \centering
 \includegraphics[scale=0.07]{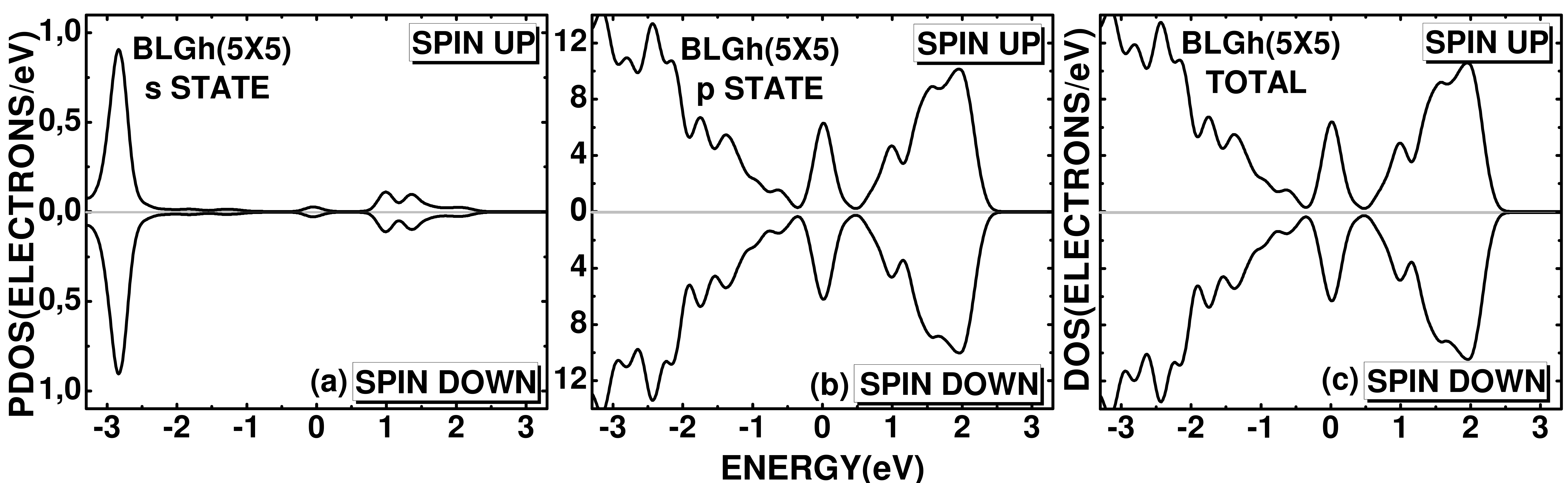}
 \caption{\label{band5x5} Spin-resolved  density of states projected on the deformed region for the 5$\times$5 structure. Note the reflection symmetry around the horizontal axis indicating that both spins are in the same state and therefore, paired. }\label{bandstrutureblg5x5}
\end{figure*}

Evidence that the zero modes are localized on the deformed region created by the intercalated atom, comes from Fig. \ref{bandstrutureblg5x5}. There, we present the spin-resolved  projected density of states (PDOS) on the site of the deformation for the $5\times$5 structure.  The   PDOS is shown separately for the $s$ and $p$ states, as well as for their sum. There is no clear spin separation in the outcome of our calculations which is natural in an all-carbon structure.  Furthermore,  the fact that both $s$ and $p$ states have peaks in the same energy region, confirms the  $sp^3$ hybridization observed in Fig. \ref{estruturas}.

The occurrence of the flat bands at zero energy in the larger periodicity BLGh is the main result of this article. We interpret this finding under the light of the pseudomagnetic field associated to distortions of graphene \cite{manes2007symmetry,guinea2010energy,levy2010strain}. The extra carbon atoms are incorporated to one of the layers more like a topological defect than as a dopant, since it is an all-carbon structure. Alternatively, the same structures can, in principle, be obtained by suitably cutting bonds, rearranging them, changing the hybridization of the involved atoms from $sp^2$ to $sp^3$  and reconnecting the appropriate bonds. So, what we have is a periodic array of ``bubbles" very much like the bubble-wrap carbon reported in \cite{liu2018bubble}. Incidentally, the band structures shown in reference \cite{liu2018bubble} present almost flat bands near the Fermi level.

%\begin{figure}[htb!]
% \centering
% \includegraphics[scale=0.25]{pictures/densityblg5x5top.png}
% \caption{\label{densityblg5x5top}}  
%\end{figure}

\section{Conclusion}

We performed first-principles calculations for a graphene bilayer periodically intercalated with  carbon atoms in the hollow position, using the DFT formalism with the GGA approach and PBE functional, implemented in the CASTEP module. In our study, we worked with four sizes of supercells, therefore increasing the periodicity of the lattice of inserted atoms. The extra atoms are found to bind covalently to one of the layers with $sp^3$ hybridization, which introduces an out-of-plane distortion. Since the inserted atoms are of the same species, their inclusion just creates a topological defect, {locally} deforming  the layer. The result is a periodic array of topologically stable deformations in one of the layers. If it were a single layer this would act as an array of  pseudomagnetic fluxes threading the structure. Since we have a bilayer, the interlayer hopping is affected by the deformations and we effectively have a non-Abelian gauge field as the one studied in \cite{san2012non}. For each of the systems generated in our study,  it was obtained the  electronic band structure and density of states. The major difference between the band structure of pristine BLG and the one with the periodic deformation in one of the layers is the appearance of two new bands spread around the Fermi level in the latter. As the period of the deformation lattice increases these two bands get narrower and narrower tending to a single degenerate flat band at the Fermi level. Exactly as predicted in Ref. \cite{san2012non}. 

{It is worth mentioning that flat, or partially flat, bands also appear in bilayer graphene  under the application of a perpendicular magnetic field \cite{le2019interplay}. Moreover, in \cite{le2019interplay} it is reported that, due to the interaction of the electron spin with the magnetic field,  the $sp^2$ hybridization is destroyed, leading to a semimetal to metal transition. In a similar way, the destruction of the $sp^2$ hybridization is  done locally  by each topological defect in our system. In this sense, our array of topological defects works very much as a lattice of magnetic flux tubes, locally changing the hybridization. The effect on the electronic structure is the opening of a small gap and the creation of flat bands in the gap. Therefore, a tight-binding calculation along the lines of Ref. \cite{le2019interplay}, for the system studied here, with and without a perpendicular magnetic field, would be very interesting for comparison with our results. Of particular interest, is the comparison between pristine BLG under a perpendicular magnetic field and  the bilayer with topological defects, in the limit of a very dense array of defects (without magnetic field).  A final comment on this process of destruction of the $sp^2$ hybridization, concerning graphene-like structures, is due here. Preliminary results from our group indicate that Bernal h-BN bilayers intercalated with  boron  atoms   also acquire flat bands in the gap, possibly due to a similar mechanism of hybridization change. This is also observed when pristine h-BN bilayer is under the action of a perpendicular magnetic field \cite{le2019interplay}. }

{Since both pristine SLG and BLG have zero band gaps, engineering both the gap and the band shape is a must for technological applications. Molecular doping \cite{samuels2013molecular}, use of an applied electric field \cite{kanayama2015gap} or even a  magnetic field \cite{le2019interplay} are known means of achieving this. Another possibility, suggested by the results presented here, is   a ``clean'' doping using carbon atoms, or  rewiring of the network, in order to create stable topological defects on the all-carbon structure. As seen above, besides opening a gap, this also introduces  flat  bands in the electronic structure,  which opens up new experimental possibilities for gap and band shape control. }

\section*{Acknowledgments}
Financial support from INCT-Nanocarbono,  Conselho Nacional de Desenvolvimento Científico e Tecnológico (CNPq), Coordenação de Aperfeiçoamento de Pessoal de Nível Superior (CAPES) and Fundação de Amparo à Ciência e Tecnologia do Estado de Pernambuco (FACEPE) is acknowledged. DLA acknowledges the Mato Grosso Research Foundation - FAPEMAT for financial support through the Grant PRONEX CNPq/FAPEMAT-850109/2009.

\section{References}
\bibliographystyle{elsarticle-num-names}
\bibliography{refs}

\end{document}